\documentclass[reprint, superscriptaddress, secnumarabic, amsmath, amssymb, nobibnotes, aps, prb]{revtex4-2}

\usepackage{graphicx}
\usepackage{caption}
\usepackage{epstopdf}
\usepackage[T1]{fontenc}
\usepackage{amsbsy}
\usepackage{gensymb}
\usepackage{amsmath}
\usepackage{amssymb}
\usepackage{bbm}
\usepackage{braket}
\usepackage{xcolor}
\usepackage{multirow}
\allowdisplaybreaks
\usepackage{graphicx}
\usepackage[hidelinks]{hyperref}  
\usepackage[normalem]{ulem}
\usepackage{orcidlink}

\renewcommand{\approx}{\simeq}

\begin{document}

	\title{\textrm{Ground-states of the Shastry-Sutherland Lattice Materials Gd$_2$Be$_2$GeO$_7$ and Dy$_2$Be$_2$GeO$_7$}}
\author{M. Pula\,\orcidlink{0000-0002-4567-5402}}
\email[]{pulam@mcmaster.ca}
\affiliation{Department of Physics and Astronomy, McMaster University, Hamilton, Ontario L8S 4M1, Canada}
\author{S. Sharma\,\orcidlink{0000-0002-4710-9615}}
\affiliation{Department of Physics and Astronomy, McMaster University, Hamilton, Ontario L8S 4M1, Canada}
\author{J. Gautreau}
\affiliation{Department of Physics and Astronomy, McMaster University, Hamilton, Ontario L8S 4M1, Canada}
\author{Sajilesh K. P.}
\affiliation{Physics Department, Technion-Israel Institute of Technology, Haifa 32000, Israel}
\author{A. Kanigel}
\affiliation{Physics Department, Technion-Israel Institute of Technology, Haifa 32000, Israel}
\author{G.~M.~Luke\,\orcidlink{0000-0003-4762-1173}}
\email[]{luke@mcmaster.ca}
\affiliation{Department of Physics and Astronomy, McMaster University, Hamilton, Ontario L8S 4M1, Canada}
\affiliation{TRIUMF, Vancouver, British Columbia V6T 2A3, Canada}		
	\date{\today}
	\begin{abstract}
		\begin{flushleft}
		\end{flushleft}
  
The recent realization that the rare-earth melilites RE$_2$Be$_2$GeO$_7$ host the Shastry-Sutherland lattice within planes of RE$^{3+}$ ions has sparked a number of studies. This family of materials lacks appreciable site mixing and conductivity, making them promising candidates for the Shastry-Sutherland model. Herein, we present the magnetic ground states of two of these rare-earth melilites: RE = Gd and Dy. We find, through measurements of magnetic susceptibility, magnetization, and specific heat capacity (RE = Dy only), that these two melilites are antiferromagnets (T$_N$ $\sim$~1~K). Gd$_2$Be$_2$GeO$_7$, in accordance with its electronic configuration, has isotropic single-ion anisotropy but shows a quadratic contribution to its magnetization. Dy$_2$Be$_2$GeO$_7$ has Ising-like single-ion ansiotropy and is likely an effective spin-$1/2$ system. Both materials exhibit metamagnetic transitions. We identify this transition in Dy$_2$Be$_2$GeO$_7$, occurring at 86(1)~mT for T=500~mK, to likely be a spin-flip transition. 
		
	\end{abstract}
	\maketitle
 
\section{Introduction}
\begin{figure*}
    \centering
    \includegraphics[width=\textwidth]{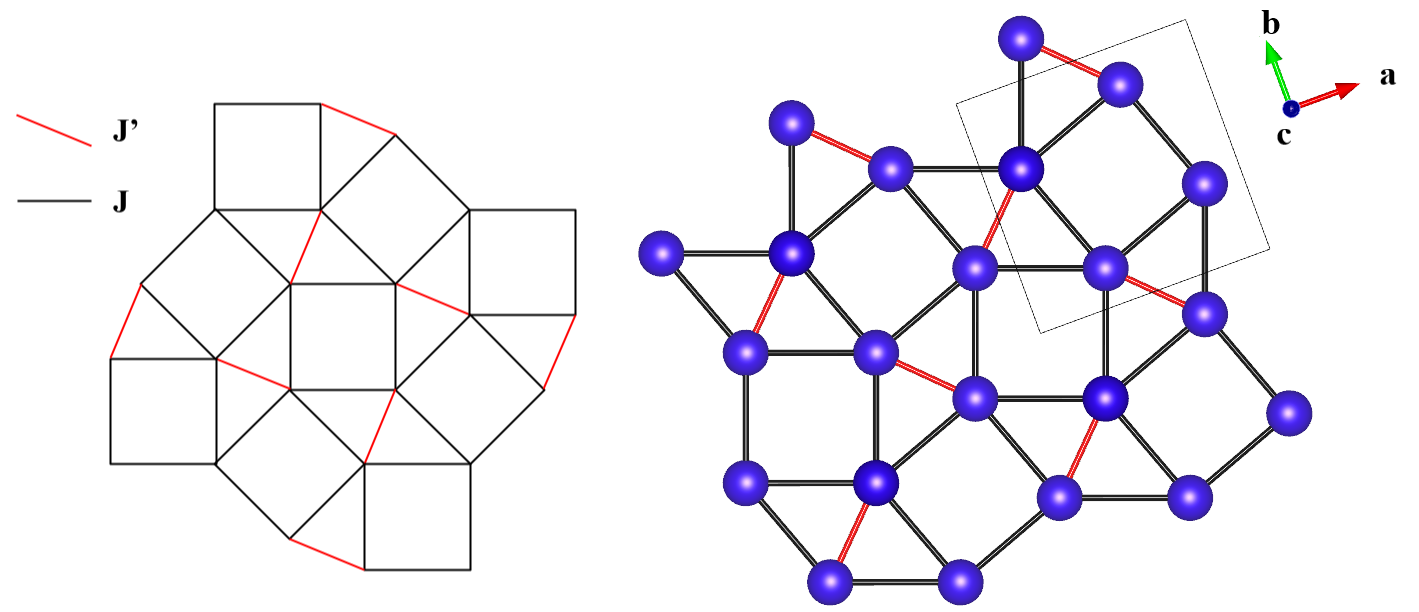}
    \caption{\textbf{Left}: The square-snub Archimedean-like Shastry-Sutherland Lattice. Here, the red bonds represent the $J'$ interactions, labeled "diagonal" in Eq. \ref{Eq:SSM}. The black bonds are the $J$ interactions, representing the "square" interactions of Eq. \ref{Eq:SSM}. \textbf{Right}:  The intraplanar crystal lattice of the rare-earth melilites Dy$_2$Be$_2$GeO$_7$ and Gd$_2$Be$_2$GeO$_7$. The unit cell is shown as the solid lines. Generated via VESTA \cite{momma2011vesta}.}
    \label{fig:Re2Be2GeO7-lattice}
\end{figure*}

The Shastry-Sutherland model is a well-known geometrically frustrated system \cite{lacroix2011introduction}. In their seminal work, Shastry and Sutherland \cite{SRIRAMSHASTRY19811069} showed that a square lattice with antiferomagnetic square nearest neighbor ($J$) and some diagonal next-nearest ($J'$) neighbor Heisenberg interactions has an exactly solvable ground state, namely a singlet dimer state ($J'~\gg~J$) or a square-lattice antiferromagnet ($J \gg J'$) \cite{SRIRAMSHASTRY19811069}. The boundary between the two phases occurs around ${J}/{J'}$ $\approx$ 0.7 \cite{miyahara1999exact}. The model is described by the Hamiltonian

\begin{equation}
    H= J \sum_{square} \mathbf{S_j} \cdot \mathbf{S_i} + J' \sum_{diagonal} \mathbf{S_j} \cdot \mathbf{S_i},
    \label{Eq:SSM}
\end{equation}
where the square and diagonal bonds are shown in Fig.~\ref{fig:Re2Be2GeO7-lattice}. 
Numerous works have followed, notably, the discovery of SrCu$_2$(BO$_3$)$_2$. SrCu$_2$(BO$_3$)$_2$ is a singlet dimer system that can be described by inter-and-intra dimer ($J$ and $J'$, respectively, in the convention used above) interactions on planes of Cu$^{2+}$ ions. This gapped system is known for being the first example of a 2-D spin system to exhibit fractional magnetization plateaus \cite{PhysRevLett.82.3168}. The intraplanar arrangement, i.e., the Shastry-Sutherland lattice (SSL), of Cu$^{2+}$ is similar to the snub-square Archimedean lattice, albeit formed with isosceles triangles rather than the required equilateral. This arrangement results in a Hamiltonian that is topologically equivalent to the Shastry-Sutherland model \cite{doi:10.1126/science.1075045, M.Albrecht_1996}.

Theoretical work has shown that additional phases, such as the plaquette \cite{koga2000quantum,corboz2013tensor} and quantum spin liquid \cite{wang2022quantum,kelecs2022rise, yang2022quantum} are also possible in the Shastry-Sutherland model, occurring for $J/J'$ ratios between the singlet and antiferromagnetic (also known as N\'eel) phases. The plaquette phase has been observed in SrCu$_2$(BO$_3$)$_2$ with the application of pressure, which tunes the $J/J'$ ratio \cite{jimenez2021quantum, guo2020quantum}.

In addition to SrCu$_2$(BO$_3$)$_2$, other SSL materials are known. These include REB$_4$ \cite{siemensmeyer2008fractional, yoshii2008multistep, ye2017electronic, michimura2006magnetic}, (Yb,Ce)$_2$Pt$_2$Pb \cite{miiller2016magnetic, kim2008yb, kabeya2018antiferromagnetic}, BaRE$_2$Zn(O,S)$_5$ \cite{marshall2023field, billingsley2022single, ishii2021magnetic,ishii2020high}, and Re$_2$Be$_2$(Ge,Si)O$_7$ \cite{brassington2024magnetic, brassington2024synthesis, pula2024candidate, yadav2024observation, liu2024distinct}. Investigation of the latter, the rare-earth (RE)-based family of melilite-like materials RE$_2$Be$_2$GeO$_7$, have recently garnered interest as SSL materials following the work of Ashtar and Bai \textit{ et al.} \cite{doi:10.1021/acs.inorgchem.0c03131}. 

Like SrCu$_2$(BO$_3$)$_2$, these rare-earth melilites form an intraplanar snub-square Archimedean-like lattice of magnetic (in the case of the melilites: RE$^{3+}$) ions. Ashtar and Bai report the magnetic properties of these melilites  down to $\sim$~2~K, but the ground-states, with the exception of RE~=~Tb, were not determined in their work. Several studies on these rare earth melilites have since emerged, including RE~=~Yb, Pr, Nd, and Er \cite{liu2024distinct, pula2024candidate,yadav2024observation}. Er$_2$Be$_2$GeO$_7$ has been shown to host long-range order and fractional magnetization plateaus \cite{yadav2024observation, pula2024fractionalized}; Nd$_2$Be$_2$GeO$_7$ has simultaneous short-range spin correlations and long-range order \cite{liu2024distinct}; Pr$_2$Be$_2$GeO$_7$ exhibits spin freezing with a lack of long-range order, possibly implying a spin-ice ground-state \cite{liu2024distinct}; and Yb$_2$Be$_2$GeO$_7$ displays persistent spin dynamics with no long-range order, potentially realizing a quantum spin liquid ground-state \cite{pula2024candidate}. 

Herein, we report the low-temperature magnetic properties of two additional rare-earth melitites: Dy$_2$Be$_2$GeO$_7$ and Gd$_2$Be$_2$GeO$_7$, which we studied through bulk SQUID magnetometry (susceptibility and magnetization) and thermodynamic measurements (specific heat capacity, RE~=~Dy only). These materials exhibit antiferromagnetic long-range order in (close to) zero-field conditions, with T$_N\sim1$~K. Both materials undergo a metamagnetic transition. For Dy$_2$Be$_2$GeO$_7$, we identify the transition (at 86(1)~mT for T~=~500~mK) as likely a spin-flip. Dy$_2$Be$_2$GeO$_7$ shows evidence of having an effective spin-$1/2$ ground-state, as the magnetic entropy in zero field nearly reaches $R\ln(2)/$Dy$^{3+}$, and the saturation magnetization implies Ising single-ion anisotropy. Conversely, Gd$_2$Be$_2$GeO$_7$ is isotropic and a quadratic magnetization component (on the order $\sim10^{-4}$~cm$^3$/[Oe~mol]) is observed in the ordered state.

\section{Methods}

Dy$_2$Be$_2$GeO$_7$ and Gd$_2$Be$_2$GeO$_7$ were synthesized in powder form through solid-state reactions. This was achieved using the techniques reported by Ashtar and Bai {\em et al.} \cite{doi:10.1021/acs.inorgchem.0c03131}, and Ochi \emph {et~al.} \cite{OCHI1982911}. (Dy/Gd)$_2$O$_3$ (4N), BeO (4N), and GeO$_2$ (4N) powders were mixed in stoichiometric ratios and subsequently ground, all while in an argon environment. The mixture was then heated in an alumina crucible at 1350~$\degree$C for a 24~h period. The grinding/heating cycle was repeated until a pure sample (2-4 cycles) was obtained.

X-ray diffraction (XRD) performed on a Panalytical X'pert Pro diffractometer was used to confirm sample composition. Isothermal magnetization and magnetic susceptibility measurements were conducted using a Quantum Design MPMS XL magnetometer for measurements in the temperature range from 1.8~K to 300~K. Low-temperature measurements were carried out with an iQuantum $^3$He insert in the temperature range from 490~mK to 6~K. Specific heat capacity measurements used the two-$\tau$ relaxation method on a Quantum Design PPMS equipped with a dilution refrigerator. La$_2$Be$_2$GeO$_7$ was used as a non-magnetic analog to remove the contribution of the lattice to the specific heat. The isothermal susceptibility $\chi(H) = dM/dH$ and the first temperature derivative of the susceptibility $d\chi(T)/dT$ are calculated as the mean slope of the two nearest adjacent points.

\section{G\MakeLowercase{d}$_2$B\MakeLowercase{e}$_2$G\MakeLowercase{e}O$_7$}
\subsection{Results}

\begin{figure*}
    \centering
    \includegraphics[width=\textwidth]{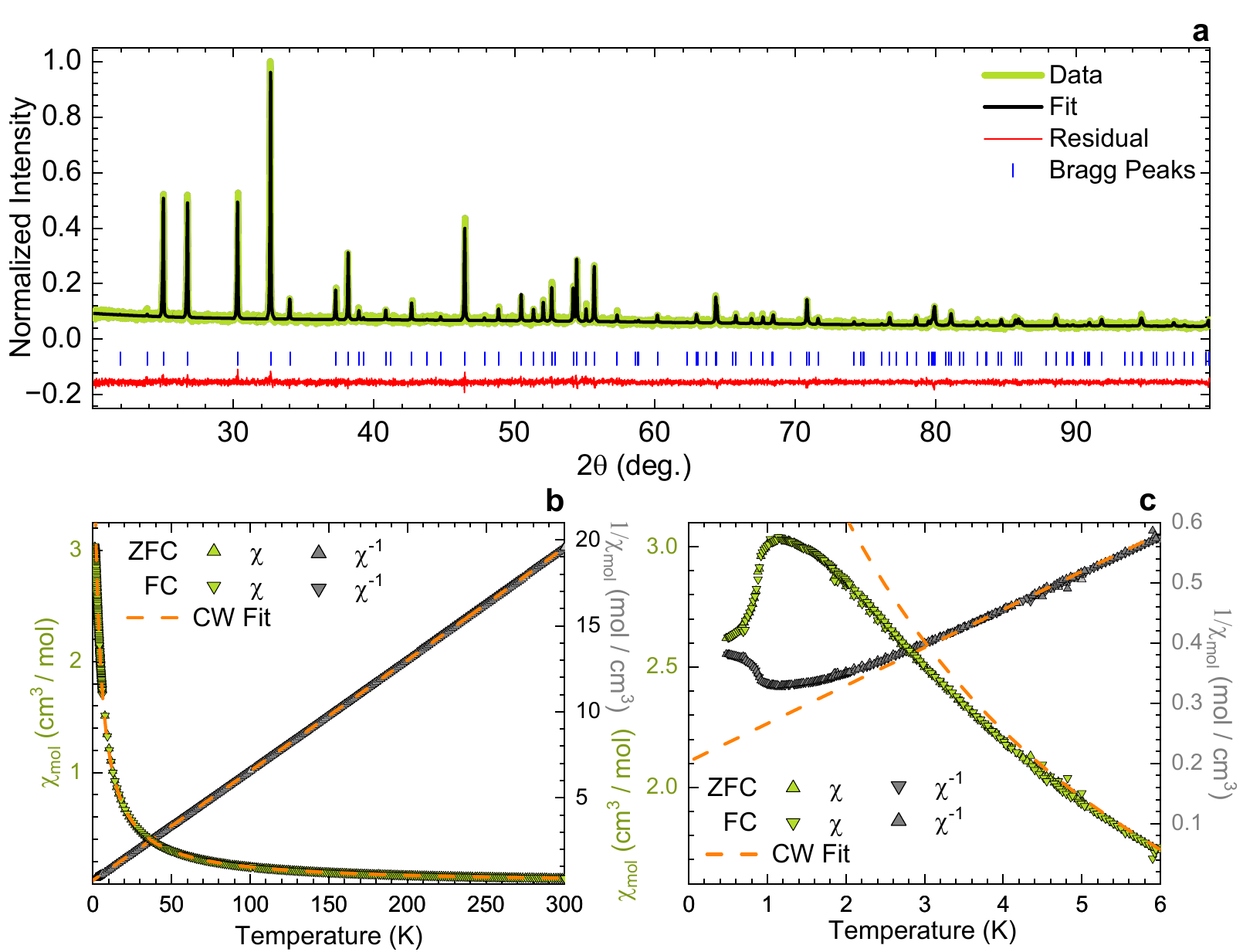}
    \caption{\textbf{Gd$_2$Be$_2$GeO$_7$} \newline 
    \textbf{a}: The X-ray diffraction pattern of Gd$_2$Be$_2$GeO$_7$. The wavelength used was 1.541 \AA. The blue tick marks indicate the Bragg peaks of the tetragonal phase Gd$_2$Be$_2$GeO$_7$ with space group P$\overline{\mbox{4}}$2$_1$m (113). The resulting patterns can be seen to be well-modeled by this phase. \textbf{b}: High-temperature $\chi$(T) (green) and 1/$\chi$(T) (grey) at 10~mT on a sample with mass 10.53~mg. ZFC measurements are represented by triangles pointing upward, while FC are pointing downward. The orange curve corresponds to the Curie-Weiss fit. \textbf{c}: Low-temperature $\chi$(T) and 1/$\chi$(T). The orange curve is again the Curie-Weiss fit. A cusp-like feature is clearly visible in the susceptibility around 1~K. The Rietveld refinement and Curie-Weiss parameters can be found in Table \ref{Tab:Gd-Dy-Paras}.}
    \label{fig:Gd-XRD-susc}
\end{figure*}

X-ray diffraction of Gd$_2$Be$_2$GeO$_7$ is presented in Fig.~\ref{fig:Gd-XRD-susc}a. The synthesized powder is single phase, with the tetragonal space group P$\overline{\mbox{4}}$2$_1$m (113). The lattice parameters, determined via Rietveld refinement, are  a~=~b~=~7.44848(4) \AA \space and c~=~4.81974(3) \AA. Additional parameters derived, such as atomic positions, can be found in Table \ref{Tab:Gd-Dy-Paras}.

 \begin{figure*}
     \centering
     \includegraphics[width=\textwidth]{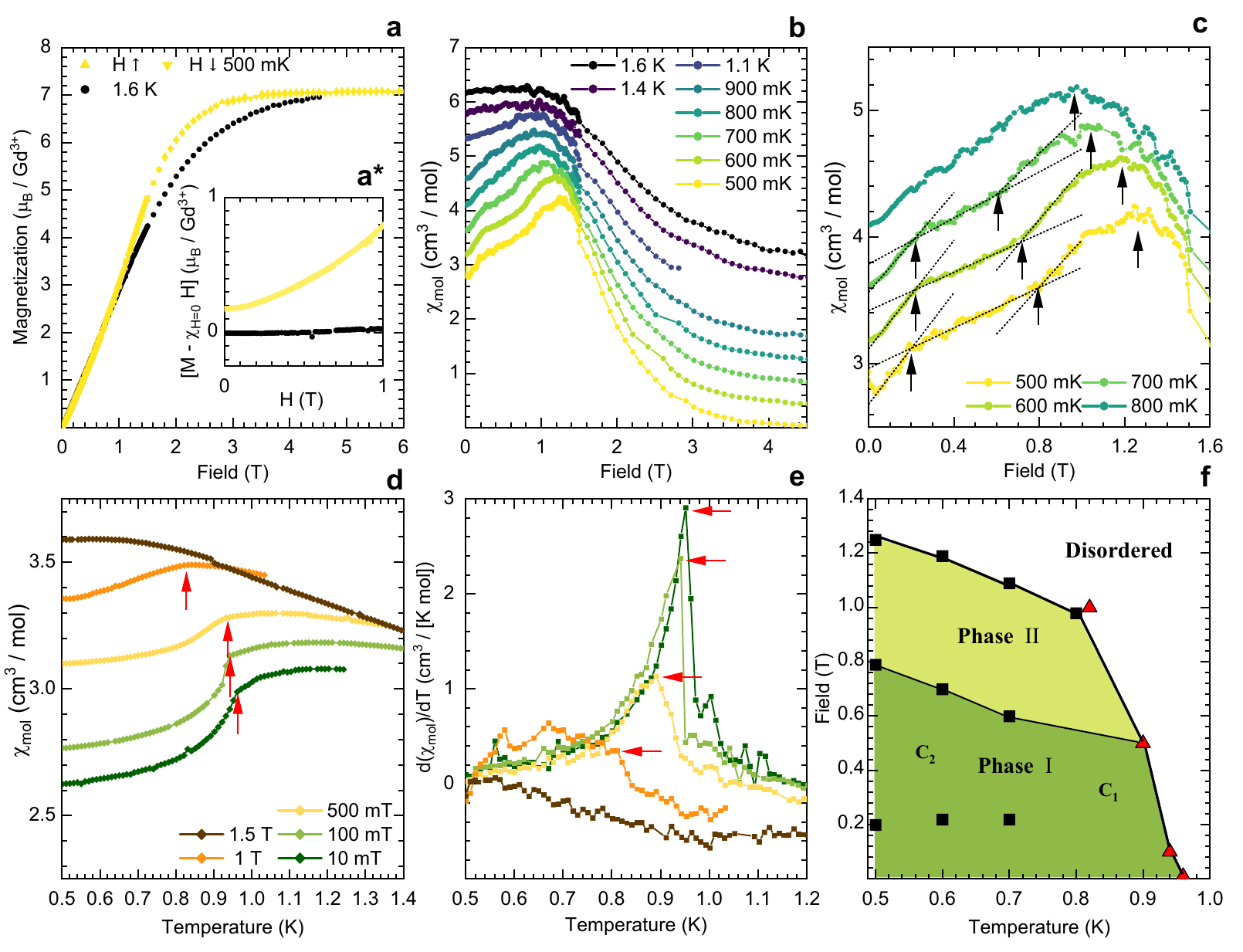}
     \caption{\textbf{Gd$_2$Be$_2$GeO$_7$} \newline All displayed measurements were performed on a sample of mass 3.5~mg. \textbf{a}: The magnetization at 500~mK and 1.6~K. \textbf{a*}: The quadratic component of the magnetization at 500~mK and 1.6~K, isolated by removing the linear contribution, i.e., $\chi^{H=0}~ H$. \textbf{b}: The isothermal susceptibility at several temperatures. The susceptibilities are offset along the y-axis for clarity. \textbf{c}: The isothermal susceptibility (again, offset along the y-axis) at low fields for a select number of temperatures. Black arrows show the deduced points used in \textbf{f}. For the points within or on the boundary of phase I, linear fits above/below each region were used to identify each point (as the data point closest to the intercept of the fits). \textbf{d}: The magnetic susceptibility at various fields. The red arrows indict the location of the phase transition for each field. \textbf{e}: The temperature derivative of $\chi$. Similar to \textbf{d}, the red arrows show the location of the phase transition (corresponding to the cusp in $\chi(T)$) at each field). \textbf{f}: An approximate phase diagram. The black squares and red triangles represent points determined using $\chi(H)$ (black arrows in \textbf{c}) and $\chi(T) + d(\chi(T)/dT$ (red arrows in \textbf{d,e}), respectively. The solid lines are guides only. The labels $C_1$ and $C_2$ represents the regions where a quadratic component of the magnetization is detected.}
     \label{fig:Gd-summary}
 \end{figure*}

 \begin{figure*}
    \centering
    \includegraphics[width=\textwidth]{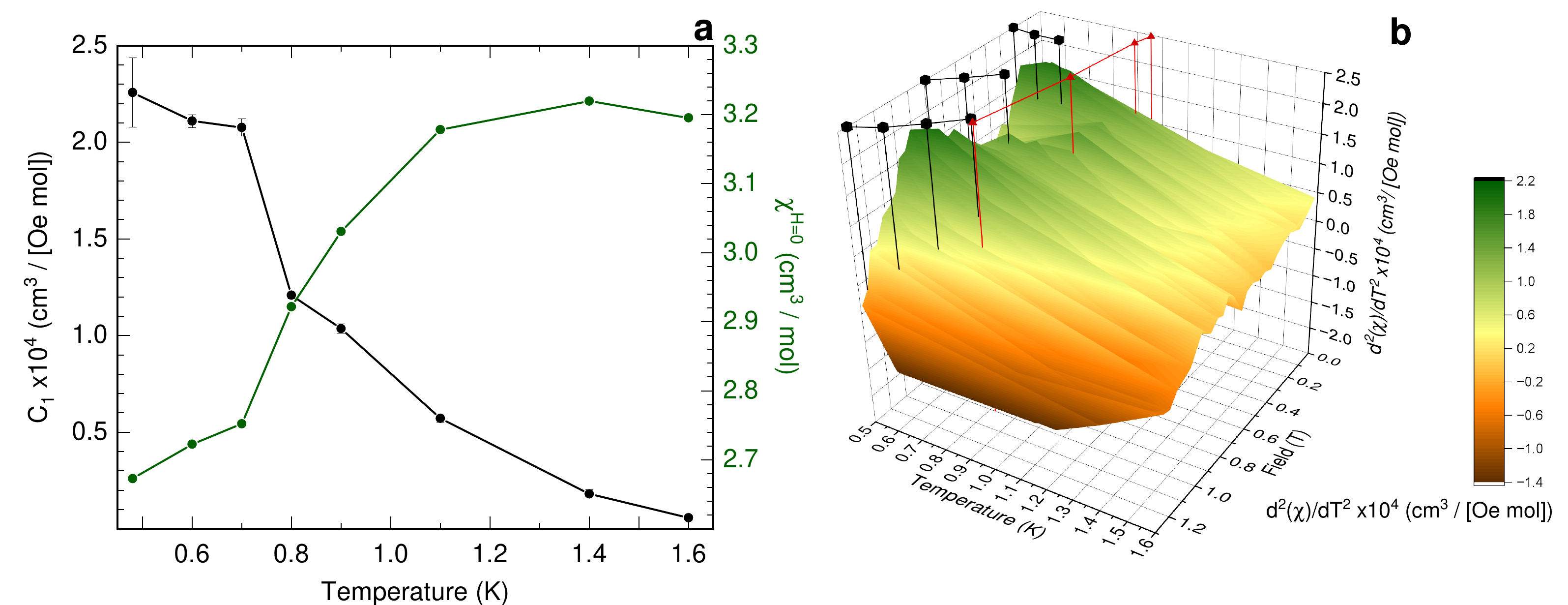}
    \caption{\textbf{Gd$_2$Be$_2$GeO$_7$} \newline 
    \textbf{a}: The temperature dependence of the coefficient $C_1$ and $\chi^{H=0}$, determined via linear fitting of $\chi(H)$. At temperatures below and including 700~mK, a fitting range of 0.05~mT to 0.195~mT was used. Above and including 800~mK, the range used was 0.10~mT to 0.73~mT. \textbf{b}: A contour plot of $d^2(\chi(T))/dT^2$, showing the value of the slope as a function of temperature and field. The black spheres and red triangles are the same points as in Fig. \ref{fig:Dy-summary}f.}
    \label{fig:Gd-para}
\end{figure*}
The magnetic susceptibility of Gd$_2$Be$_2$GeO$_7$ (see Fig.~\ref{fig:Gd-XRD-susc}b for the entire temperature range and Fig.~\ref{fig:Gd-XRD-susc}c for low temperatures) is well modeled across almost (300~K~$\rightarrow$~5~K) the entire temperature range by the Curie-Weiss law with an additional term
\begin{equation}
    \chi = \frac{C}{T -\theta} + \chi_0
    \label{eqn:CWLaw}
\end{equation}
where $\chi_0$ is a temperature independent susceptibility term, which was used to remove the diamagnetic contribution. Using this model in a fitting range of 300~K~$\rightarrow$~20~K results in an effective moment of 7.92(1)~$\mu_B/$Gd$^{3+}$ (very close to the free-ion moment of $\mu_J$~=~7.94~$\mu_B/$Gd$^{3+}$) and a Weiss temperature of $\theta=-~3.05(3)$~K. The deduced Weiss temperature is negative, implying that the predominant magnetic interactions are antiferromagnetic. The behavior seen in the low-temperature magnetic susceptibility, namely a hysteresis-free cusp-like feature around 1~K, is reminiscent of an antiferromagnetic (AFM) transition (see Fig. \ref{fig:Gd-XRD-susc}c). The fitting parameters are summarized in Table \ref{Tab:Gd-Dy-Paras}. Our results are largely in agreement with Ashtar and Bai's results \cite{doi:10.1021/acs.inorgchem.0c03131}, although our susceptibility parameters vary somewhat due to a difference in modeling (Ashtar and Bai employ a two-slope approach, i.e., individual high- and low-temperature Curie-Weiss fits). 

 No evidence of crystal electric fields are observed in the susceptibility, evidenced by a lack of slope change in the inverse susceptibility down to Kelvin-scale temperatures (see Fig. \ref{fig:Gd-XRD-susc}b). This is expected because the electronic configuration of Gd$^3+$, namely 4f$^7$, having L~=~0, precludes orbital angular momentum contributions. The ground state is therefore concluded to be an isotropic antiferromagnet. The slight curvature near T$_N$ is evidence of the breakdown of the Curie-Weiss law as the magnetic interactions become relevant compared to the temperature scale. 

 The isothermal magnetization can be seen in Fig.~\ref{fig:Gd-summary}a. The magnetization saturates to $\sim$~7.1(1)~$\mu_B/$Gd$^{3+}$ 
 by $\sim$ 3~T, in good agreement with the expected $M_S~=~gS\mu_B~=~7\mu_B$. Interestingly, for temperatures below $\sim$~1.4~K, the isothermal susceptibility has at least one region of appreciable linear increase with field (see Figs.~\ref{fig:Gd-summary}b and \ref{fig:Gd-summary}c). Equivalently, the magnetization increases quadratically with field (see Fig. \ref{fig:Gd-summary}a*). Such an effect has been observed in, e.g., the rare-earth othroferrites (REFeO$_3$) \cite{gorodetsky1964second, gorodetsky1967linear} and is allowed for any magnetic point group that does not have time-inversion symmetry \cite{kharchenko1995quadratic, gorodetsky1967linear}, implying a noncollinear spin structure. The magnetic susceptibility in this case is \cite{gorodetsky1967linear}
\begin{equation}
    \chi (H) = \chi^{H=0} + C~H,
    \label{Eq:chi}
\end{equation}
where $\chi^{H=0}$ and $C$ correspond to the coefficients of the quadratic and cubic terms in a power-law expansion of the magnetic energy.

Below temperatures of $\sim700$~mK, four distinct regions can be seen in the isothermal susceptibility (see Fig. \ref{fig:Gd-summary}b and \ref{fig:Gd-summary}c). There are two regions which are linear in field, a "hump"-like region, and a subsequent region that appears Brillouin-like, i.e., disordered. The linear regions are characterized by slopes labeled $C_1$ (the first region, that is, lower fields) and $C_2$ (the second region, i.e., higher fields). The range of these regions was determined by the intersection of linear fits above and below each region (see Fig. \ref{fig:Gd-summary}c). The temperature dependence of $C_1$ (shown in Fig. \ref{fig:Gd-para}a) decreases monotonically, reaching essentially zero by 1.6~K. $C_2$ was found to be roughly temperature independent and equal to 8.4(7)$\times 10^{-5}$~cm$^3/$[Oe~mol]. $\chi^{H=0}$ was determined as the intercept of the linear fit in the first region (that is, of $C_1$) (see Fig. \ref{fig:Gd-para}a). Above the second linear regime (e.g., above $\sim$~790(10)~mT at 500~mK), a broad maximum is observed. The sudden change in behavior of the isothermal susceptibility implies phase transitions, and hence we take the onset of this maximum to indicate a metamagnetic transition into a phase we label "phase II". Subsequently, this transition is followed by the disordered/spin-polarized state for sufficiently large fields (1.25(1)~T at 500~mK). Magnetic susceptibility measurements confirm (see Fig. \ref{fig:Gd-summary}d) the lack of a phase transition at any measured temperature in a field of 1.5~T, evidenced by the lack of a cusp in susceptibility and corresponding peak in $d\chi (T)/dT$. This implies that the phase remains disordered above fields of 1.5~T (isothermal magnetization narrows this to 1.25(1)~T).

For temperatures above and including 800~mK, the linear and "hump" regions are indiscernible [see Fig. \ref{fig:Gd-para}b for a contour plot of the slope of $d(\chi(T))/dT$)]. The second linear region is also absent, leaving only one $C$ coefficient, which we label $C_1$. The temperature dependencies of $C_1$ and $\chi^{H=0}$ are shown in Fig.~\ref{fig:Gd-para}a. A diagram of the approximate phase diagram is shown in Fig. \ref{fig:Gd-summary}f. The boundary of the disordered phase is determined from both $\chi(H)$ (as the onset of a Brillouin-like behavior) and $\chi(T)$ [as the locations of the cusp, corresponding to a peak in $d\chi (T)/dT$)]. The boundary between phase I and phase II is taken to be the onset of the "hump" in $\chi(H)$. For the 1~T data set, which lacks a sharp peak, the shoulder of the broad maximum nearest the inflection point is taken as the entry from disorder into phase II. 

\begin{figure*}
    \centering
    \includegraphics[width=\textwidth]{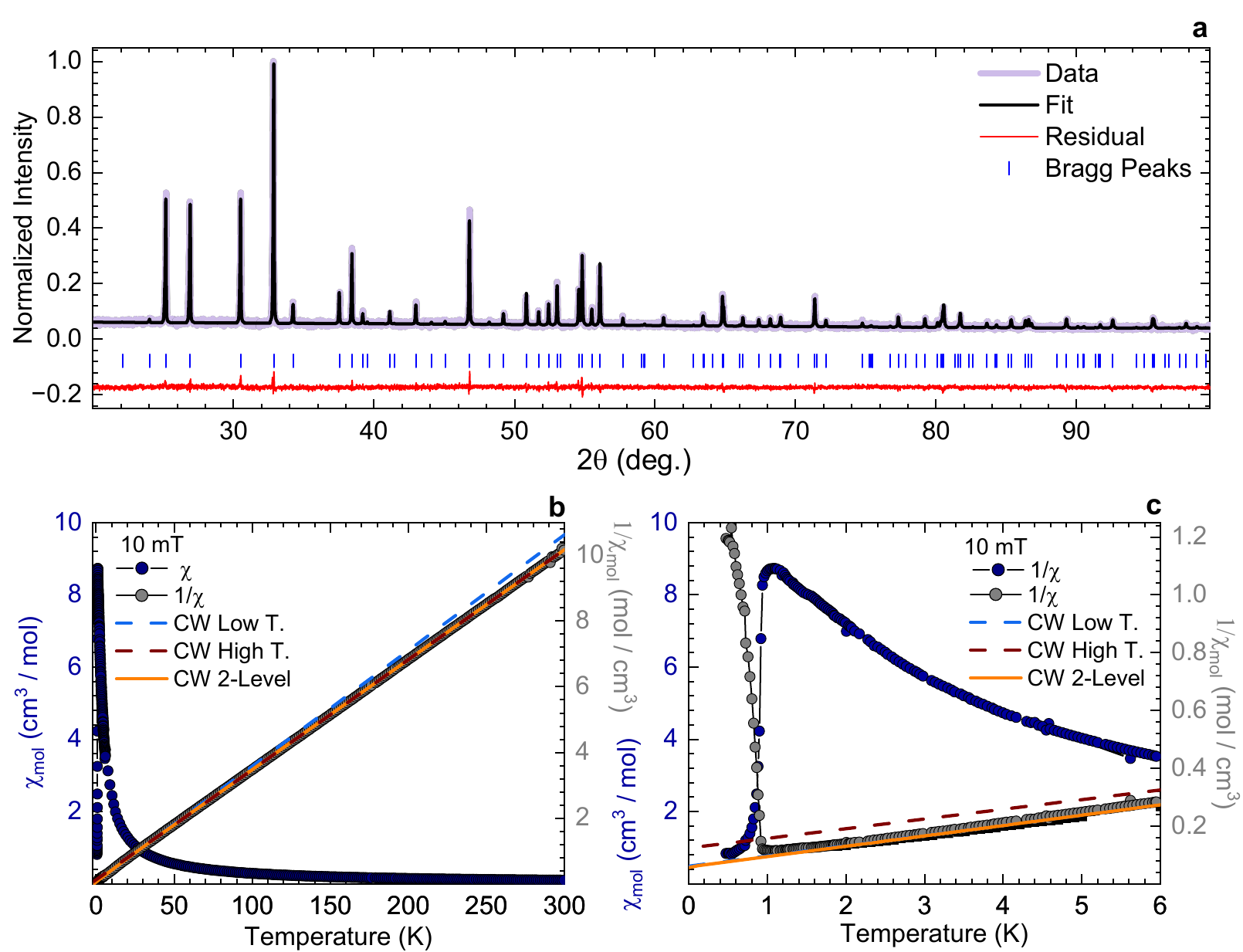}
    \caption{\textbf{Dy$_2$Be$_2$GeO$_7$} \newline 
    \textbf{a}: The X-ray diffraction pattern of Dy$_2$Be$_2$GeO$_7$. The wavelength used was 1.541 \AA. The blue tick marks correspond to the Bragg peaks associated of the tetragonal phase Dy$_2$Be$_2$GeO$_7$ with space group P$\overline{\mbox{4}}$2$_1$m (113). The resulting patterns can be seen to be well-modeled by this phase. \textbf{b}: The magnetic susceptibility and inverse susceptibility with single and two-level Curie-Weiss (CW) fittings. The blue/red dashed line is the low/high temperature Curie-Weiss fit, while the orange solid line is the two-level fit. The sample mass is 20.20~mg. \textbf{c}: The magnetic susceptibility and inverse susceptibility at low temperatures. Note the cusp near $\sim$~1~K, and the magnitude of the susceptibility below $T_N$ approaches zero.}
    \label{fig:Dy-XRD-susc}
\end{figure*}

\subsection{Discussion}

Gd$_2$Be$_2$GeO$_7$ has the signatures of an isotropic antiferromagnet (Phase I in Fig. \ref{fig:Gd-summary}f). The negative Weiss temperature ($\theta=-3.05(3)$~K) indicates that magnetic interactions are predominantly antiferromagnetic. The magnetic susceptibility displays a cusp around 1~K. The magnetization saturates to 7.1(1) $\mu_B/$Gd$^{3+}$, in agreement with the expected 7$\mu_B/$Gd$^{3+}$ in the absence of magnetocrystalline anisotropy. The Ne\'el temperature in a field of 10~mT is 960(10)~mK, a factor of $\sim$~3 times smaller than the Weiss temperature. The discrepancy between the observed Ne\'el  and Weiss temperatures indicates Gd$_2$Be$_2$GeO$_7$ is frustrated, which is expected for the SSL when the nearest-neighbor and next-nearest-neighbor interactions are both antiferromagnetic. Dipole-dipole interactions, estimated to be 600(1)~mK (square bond) and 981(1)~mK (diagonal bond) \cite{doi:10.1021/acs.inorgchem.0c03131}, are relevant on this temperature scale. Dipole-dipole interactions are known to be a source of anisotropy \cite{rotter2003dipole}, and have been proposed to be responsible for magnetic order in another Gd$^{3+}$ system: Gd$_2$Ti$_2$O$_7$ \cite{palmer2000order}. It is likely they play a role in determining the ground-state in Gd$_2$Be$_2$GeO$_7$ as well. Naively, examining a single Gd$^{3+}$ ion, which has four square and one diagonal bond, a parallel dipole-dipole interaction out-of-plane, i.e., along the c-axis, would simply sum (hence $\bf{S}\cdot \bf{r}$~$\approx$~0 if \textbf{S} is constrained to be out-of-plane) to 4~$\times$~600~mK~+~981~mK~$\sim$~3.4~K. The magnetic moment at the transition to disorder, i.e., 1.25(1)~T at 500~mK, is about 4$~\mu_B$, giving a magnetic energy of 4~$\times$~0.67~J$/$T~$\times$~1.25~T~$\sim$~3.4~K. Similarly, at the metamagnetic transition that occurs around 800~mT, the magnetic moment in this field is about $\sim$~2.3~$\mu_B$, resulting in a magnetic energy of $\sim$~1.2~K. This is about the energy of a parallel nearest-neighbor (diagonal), antiparallel next-nearest-neighbor (square) spin arrangement, which would be the difference of 4~$\times$~600~mK~-~981~mK~$\sim$~1.4~K. Although a very simplistic picture, this suggests the metamagnetic transition into phase II may be a spin-flip from a phase with competing parallel nearest-neighbor and antiparallel next-nearest-neighbor dipole interaction (antiferromagnetic) to one with all parallel arrangements (spin-flip).

\begin{figure*}
    \centering
    \includegraphics[width=\textwidth]{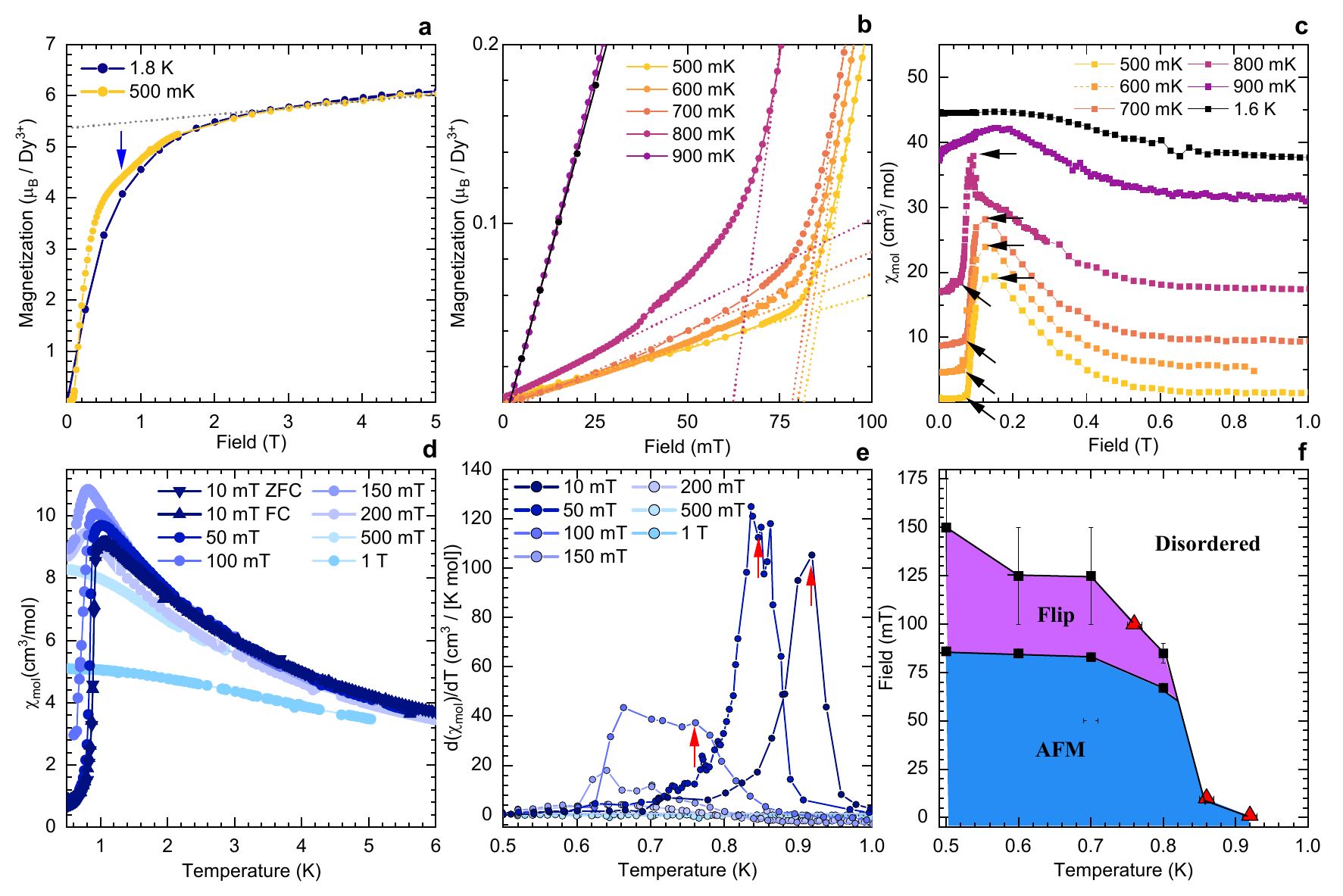}
    \caption{\textbf{Dy$_2$Be$_2$GeO$_7$} \newline \textbf{a}: The magnetization at 500~mK and 1.8~K on a sample of mass 2.8~mg. The gray solid line represents a linear fit from 3~T to 7~T, from which a saturation magnetization of 5.4(1)~$\mu_B/$Dy$^{3+}$ is predicted. The blue arrow marks the location of a likely hard-axes metamagnetic transition. \textbf{b}: The low-field magnetization at various temperatures. The intercept of the dotted lines are the critical fields at each temperature. \textbf{c}: $\chi(H)$ at various temperatures. The vertical axis is offset for clarity. The diagonal black arrows show the locations of the metamagnetic transition, while horizontal arrows show the onset of disorder. \textbf{d}: $\chi$(T) at low-temperature on a sample of mass 20.20~mg. \textbf{e}: $d\chi (T)/dT$ around the transition. The red arrows indicate the location of the phase transition at each field. \textbf{f}: The phase diagram. The black square and red triangles marks represent points deduced from $M(H) + \chi(H)$ (dotted lines in \textbf{b}, black arrows in \textbf{c}) and $d(\chi (T))/dT$ (red arrows in \textbf{e}), respectively. The solid lines are a guide only.}
    \label{fig:Dy-summary}
\end{figure*}

A quadratic field dependence to the magnetization is detected in phase I (antiferromagnetic phase). This is different from other Gd$^{3+}$ compounds, such as Gd$_2$Sn$_2$O$_7$ \cite{freitas2011magnetic}, the aforementioned Gd$_2$Ti$_2$O$_7$ \cite{raju1999transition}, and Gd$_3$Ga$_5$O$_{12}$ \cite{schiffer1994investigation}, for which no such quadratic component has been reported (to our knowledge). This quadratic dependence persists until a metamagnetic transition occurs (around 800~mT at 500~mK). At temperatures less than or equal to 700~mK, the quadratic component is characterized by two unique coefficients, $C_1$ and $C_2$. It is unclear whether the two observed quadratic components result from two distinct phases, as no evidence of a corresponding transition is apparent in $\chi(T)$. A quadratic component is allowed when the magnetic point group breaks the time-inversion symmetry, implying a non-collinear spin structure. Due to the non-centrosymmetric nature of the crystallographic space group, the Dyzloshinskii-Moriya interaction is allowed \cite{liu2024theory}. This is a possible cause of the canting of the spins, although dipole-dipole induced anisotropy is also a possibility. Thermodynamic and neutron diffraction measurements (albeit this would require Gd isotopic substitution) could be used to identify any phase transitions and investigate the magnetic structure.

\section{D\MakeLowercase{y}$_2$B\MakeLowercase{e}$_2$G\MakeLowercase{e}O$_7$}

\subsection{Results}

 The XRD pattern of the synthesized Dy$_2$Be$_2$GeO$_7$ is shown in Fig.~\ref{fig:Dy-XRD-susc}a. All Bragg peaks present in the pattern are accounted for by the tetragonal phase Dy$_2$Be$_2$GeO$_7$ with space group P$\overline{\mbox{4}}$2$_1$m (113), indicating no impurity phases are detectable in our sample. Rietveld refinement of the resulting pattern gives lattice parameters of a~=~b~=~7.41684(3)~\AA~and c~=~4.79985(3)~\AA. The deduced structural parameters are summarized in Table \ref{Tab:Gd-Dy-Paras}.

\begin{figure*}
    \centering
    \includegraphics[width=\textwidth]{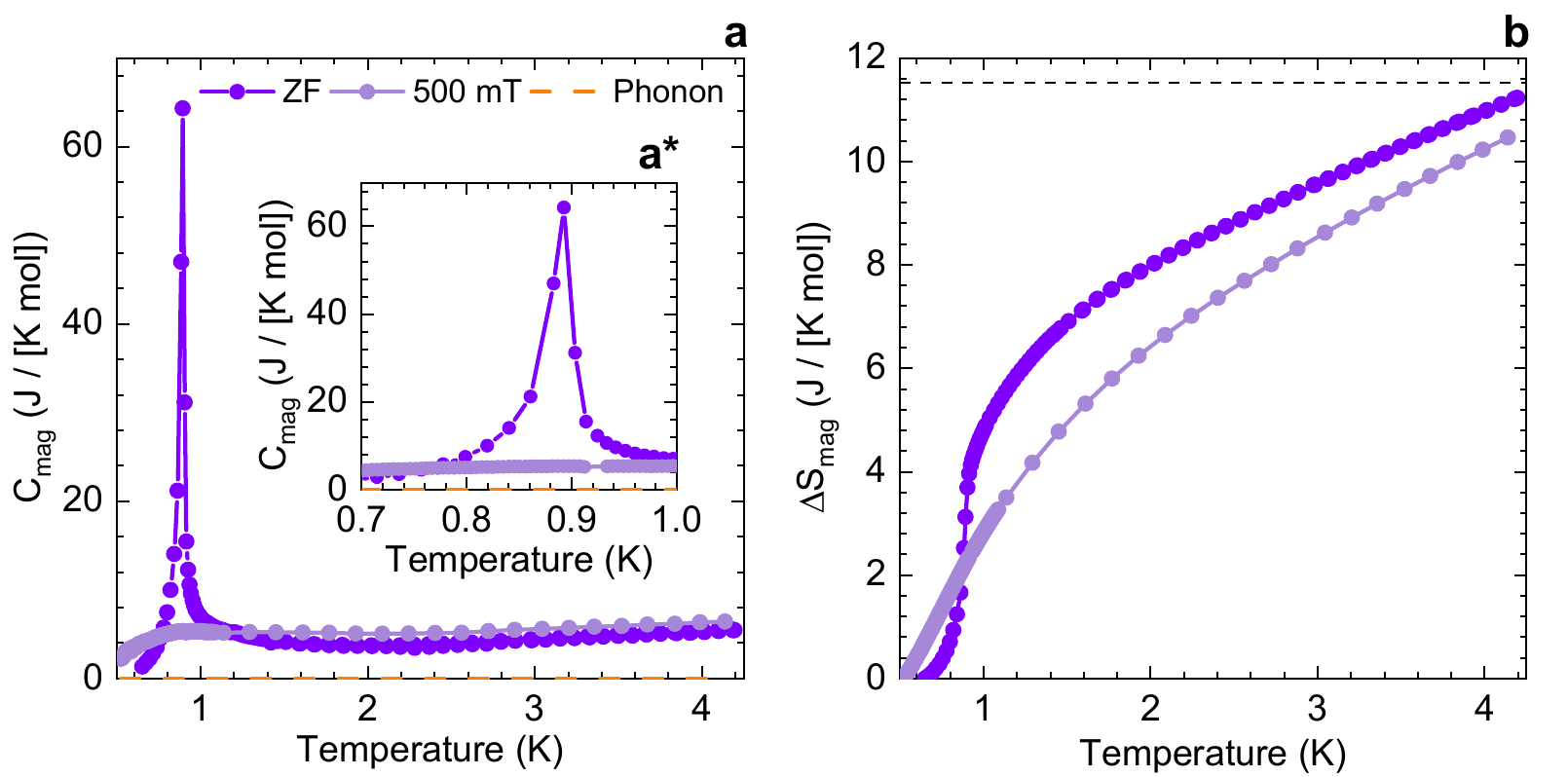}
    \caption{\textbf{Dy$_2$Be$_2$GeO$_7$} \newline
    \textbf{a}: The specific heat capacity of Dy$_2$Be$_2$GeO$_7$ in both zero and 500~mT applied fields. The solid black line in the ZF data is a visual guide. La$_2$Be$_2$GeO$_7$ was used as a non-magnetic analog to remove the phonon contribution (dashed orange line on x-axis). Of note is the complete suppression of the phase transition observed in ZF conditions under the 500~mT field. \textbf{a*}: The specific heat around the phase transition. The labels are the same as in \textbf{a}. \textbf{b}: The magnetic entropy in both zero and 500~mT fields. Again, the labels are shown in \textbf{a} (the phonon contribution is excluded). The dashed line represents an entropy of $R\ln(2)$. The magnetic entropy was calculated as $\Delta S_{mag}=\int \frac{C_{mag}}{T} dT$. }
    \label{fig:DyHeatCap}
\end{figure*}

 Magnetic susceptibility at high and low temperatures is shown in Figs. \ref{fig:Dy-XRD-susc}b and \ref{fig:Dy-XRD-susc}c. Due to the presence of crystal electric fields, a two-level phenomenological model is required to capture the behavior of the magnetic susceptibility in Dy$_2$Be$_2$GeO$_7$. This model has the form \cite{mugiraneza2022tutorial}: 
 \begin{equation}
    \frac{1}{\chi} = \frac{3 k_B}{N_A}  (T - \theta) \frac{(1 + e^{-\frac{\Delta}{k_B T}})}{\mu_{1}^2 + \mu_{2}^2~e^{-\frac{\Delta}{k_B T}}}
    \label{eqn:CWLawTwoLevel}
\end{equation}
with $\mu_1$ and $\mu_2$ being the moments of each level and $\Delta$ being the energy gap between levels. The fitting parameters of the two-level Curie-Weiss law, as well as one-level fits at high and low temperatures, are presented in Table \ref{Tab:Gd-Dy-Paras}. The effective moments of the one-level high temperature fit and two-level fits (10.890(3)~$\mu_B$/Dy$^{3+}$ and 10.9(1)~$\mu_B$/Dy$^{3+}$, respectively) are in agreement, although both slightly exceed the free-ion moment of 10.65~$\mu_B$/Dy$^{3+}$. The low temperature one-level fit and the effective moment of the two-level fit at low temperatures (which is simply $\mu_1$ in the limit of $\Delta$~$\gg$k$_b$T) are also in agreement with 10.62(2)~$\mu_B$/Dy$^{3+}$ and 10.57(3)~$\mu_B$/Dy$^{3+}$, respectively, and relatively close to the free ion moment. This is contrary to the values determined by Ashtar and Bai \cite{doi:10.1021/acs.inorgchem.0c03131}, who report an effective moment of 9.92(1) $\mu_B$/Dy$^{3+}$. The observed saturation magnetization, found to be 5.4(1)~$\mu_B$, is significantly lower than the expected saturation magnetization of $g_JJ~\mu_B$~=~10~$\mu_B$ per Dy$^{3+}$ (see Fig. \ref{fig:Dy-summary}a). This is indicative of significant anisotropy and, being about half the expected moment in a powder sample, specifically Ising anisotropy \cite{bramwell2000bulk}.

\begin{table*}
\begin{ruledtabular}
    \begin{tabular}{c|c|cccc|cccc}
    Dy$_2$Be$_2$GeO$_7$\tabularnewline
    \hline
     XRD (300~K) &Atom& x & y& z& &a (\AA)&b (\AA)& c (\AA) \tabularnewline
     $\lambda$ = 1.541 \AA&Dy & 0.1592(2) & 0.3408(2) & 0.5086(8) & &7.41684(3) &7.41684(3)& 4.79985(3)\tabularnewline
     &Be & 0.134(4)& 0.634(4) & 0.055(8) &  \tabularnewline
     &Ge & 0.0 & 0.0 & 0.0 & &  & & \tabularnewline
     &O$_1$ & 0.084(2) & 0.841(2) & 0.200(3) & & & & \tabularnewline 
     &O$_2$ & 0.144(2) & 0.644(2) & 0.737(3) & &R$_p$ &R$_{wp}$ & R$_e$ & $\chi^2$\tabularnewline
     &O$_3$ & 0 & 0.5 & 0.192(5) & & 29.0 & 21.8& 19.7& 1.233  \tabularnewline
    \hline Magnetometry && $\theta$ (K) & $\mu_{eff}$ ($\mu_B$) & $\mu_{1}$ ($\mu_B$) & $\mu_{2}$ ($\mu_B$)  & Range (K) & $\Delta$ (K) \tabularnewline
      &High T.& -3.69(6) & 10.890(4) &&& 40 $\rightarrow$ 300 \tabularnewline
      &Low T.& -1.74(3) & 10.62(2) &&& 1.89 $\rightarrow$ 40 \tabularnewline
      &Two-Level& -1.63(6) & 10.9(1) & 10.57(3) & 11.27(3)  & 1.89 $\rightarrow$ 300 & 67(6)\tabularnewline
     \hline 
     \hline

    Gd$_2$Be$_2$GeO$_7$ \tabularnewline
    \hline
    XRD (300~K) &Atom& x & y& z& &a (\AA)&b (\AA)& c (\AA) \tabularnewline
     $\lambda$ = 1.541 \AA&Gd & 0.1590(2) & 0.3410(2) & 0.5113(9) & &7.44848(4) &7.44848(4)& 4.81974(3)\tabularnewline
     &Be & 0.134(4)& 0.634(4) & 0.03(1) &  \tabularnewline
     &Ge & 0.0 & 0.0 & 0.0 & &  & & \tabularnewline
     &O$_1$ & 0.085(2) & 0.864(3) & 0.193(3) & & & & \tabularnewline 
     &O$_2$ & 0.138(3) & 0.638(3) & 0.754(4) & &R$_p$ &R$_{wp}$ & R$_e$ & $\chi^2$\tabularnewline
     &O$_3$ & 0 & 0.5 & 0.195(6) & & 41.0 & 28.1& 27.0& 1.077  \tabularnewline
      \hline Magnetometry && $\theta$ (K) & $\mu_{eff}$ ($\mu_B$) &  &&Range (K)&&&\tabularnewline
      && -3.05(3) & 7.92(1) &  &&20 $\rightarrow$ 300 \tabularnewline
    \end{tabular}
\end{ruledtabular}
\caption{A summary of the fitting parameters deduced from the structural and magnetic susceptibility analyses for Dy$_2$Be$_2$GeO$_7$ and Gd$_2$Be$_2$GeO$_7$. The effective moment of the two-level fit is calculated as the square average of $\mu_1$ and $\mu_2$. The reported errors of the structural parameters are generated by the Reitveld refinement.}
\label{Tab:Gd-Dy-Paras}
\end{table*}

The salient magnetic characteristic observed in Dy$_2$Be$_2$GeO$_7$ is a nearly zero and constant susceptibility at low temperatures and fields (see Figs. \ref{fig:Dy-summary}b and \ref{fig:Dy-summary}d), or analogously a linear magnetization with a small slope $dM/dH$. The exhibited behavior is consistent with that of an Ising antiferromagnet. The phase transition from the paramagnetic state is marked by a cusp in the susceptibility and a corresponding peak in $d(\chi(T))/dT$ (red arrows in Fig. \ref{fig:Dy-summary}e). This is followed by a metamagnetic transition for a sufficiently large magnetic field. Considering that the saturation moment indicates that Ising anisotropy is present in applied magnetic fields of at least the Tesla scale, the observed transition is likely a spin-flip. The critical field is deduced by assuming that the magnetization is linear above and below the critical field and extrapolating the intercept (see Fig. \ref{fig:Dy-summary}b). The exact fields at which these transitions occur are temperature dependent (see Fig. \ref{fig:Dy-summary}b and Fig. \ref{fig:Dy-summary}f). For a temperature of 500~mK, this transition occurs for a field of 86(1)~mT. A critical field of 86(1)~mT is consistent with the magnetic susceptibility, for which no sharp peaks in $d\chi (T)/dT$ are observed in fields $\ge$ 100~mT (see Fig. \ref{fig:Dy-summary}e), but which are present at 100~mT and below. Like Gd$_2$Be$_2$GeO$_7$, a broad maximum is observed, and we consider the shoulder closest to the inflection point to be the transition into the spin-flip phase. The isothermal susceptibility above this field increases dramatically, reaching a peak at 150(25)~mT at 500~mK (horizontal black arrows in Fig. \ref{fig:Dy-summary}c). Following this, Brillouin function-like behavior is observed. We therefore take the local maximum of the peak of $\chi(H)$ to be the onset of paramagnetism/spin polarization. Note that the field-induced metamagnetic transition is expected to be different for the easy-axis and hard-axis in the presence of magnetocrystalline anisotropy. There is evidence of an additional transition in the isothermal magnetization in the vicinity of $\sim800$~mT at 500~mK (blue arrow in Fig. \ref{fig:Dy-summary}a). This is likely a metamagnetic transition associated with the hard axis (hence it occurs at larger fields), as seen, for instance, in single crystals of Dy$_2$Ge$_2$O$_7$ \cite{ke2008magnetothermodynamics}. However, the effect is subtle, and hence the temperature dependence cannot be accurately determined in our data set. Figure \ref{fig:Dy-summary}f contains a phase diagram; the boundaries are established using $M(H) + \chi(H)$ (black arrows in Fig \ref{fig:Dy-summary}c), and $d\chi (T)/dT$ (red arrows in \ref{fig:Dy-summary}e) as described above.

To eliminate the possibility of a singlet dimer ground state, which would also have a zero-and-constant susceptibility in the singlet state, we performed thermodynamic measurements. In zero field, the magnetic specific heat capacity shows a clear $\lambda$-like peak centered at 900~mK (see Fig. \ref{fig:DyHeatCap}a*),  indicative of a phase transition and thus confirming long-range order. The heat capacity remains finite well above this transition, with a long tail that extends to the highest measured temperature of 4.2~K (see Fig. \ref{fig:DyHeatCap}a). The calculated entropy through the entire temperature range reaches about $R\ln(2)$, indicative of a ground state with two microstates (see Fig. \ref{fig:DyHeatCap}b), i.e., an effective spin-$1/2$ system. However, the specific heat capacity can be seen to increase with temperature past $\sim$~2.5~K, implying the full magnetic entropy may exceed $R\ln(2)$. Interestingly, in a 500~mT applied field, the $\lambda$-like peak is entirely suppressed (see Fig.~\ref{fig:DyHeatCap}a), implying a lack of long-range order. Instead, the heat capacity at temperatures above and below the ZF T$_N$ increases, suggesting that short-range correlations are enhanced with applied field. A slight decrease in the entropy reached at 4.2~K is likely a consequence of the temperature range measured, specifically the lower 500~mK limit (where there is still significant heat capacity in the 500~mT data).

\subsection{Discussion}

Having a nearly zero and constant susceptibility implies a ground state that is strongly pinned by anisotropy or that is nonmagnetic (a dimer singlet). Given a phase transition is observed in the specific heat capacity (which is not typically expected for a dimer singlet ground state) and a magnetization that saturates to about half the expected moment (the powder average saturation magnetization of an Ising AFM), it is likely Dy$_2$Be$_2$GeO$_7$ is an Ising antiferromagnet. The metamagnetic transition that occurs around 86(1)~mT (at 500~mK) can be attributed to a spin-flip transition. The lack of ordering in a 500~mT field, coupled with a saturation magnetization that reaches $\sim~1/2$ the expected full moment (even at temperatures above T$_N$) implies that the single-ion effects remain intact, but the two-ion effects, i.e., the antiferromagnetic interaction, are broken by a relatively small field. Interestingly, Dy$_2$Be$_2$GeO$_7$ is remarkably similar to the tetragonal pyrogermanate Dy$_2$Ge$_2$O$_7$ \cite{ke2008magnetothermodynamics}, which has been shown \cite{zhou2011high, zhou2012chemical} to be a spin ice when synthesized under high-pressure conditions (giving a pyrochlore structure). Both materials exhibit a region of small, constant susceptibility below a critical field, above which a metamagnetic transition occurs and long-range order is suppressed. Unlike Dy$_2$Ge$_2$O$_7$,  Dy$_2$Be$_2$GeO$_7$ appears to have no zero-field residual entropy, and the suppression of the long-range order appears far more complete at an equivalent field of 500~mT, despite having a larger critical field. The specific heat capacity of Dy$_2$Be$_2$GeO$_7$ is almost entirely constant throughout the measured temperature range of 500~mK to 4.2~K in a 500~mT field, having no resolvable features throughout the measured temperature range (excluding decreasing towards zero for low temperatures). A neutron study (like Gd, this would be complicated by the strong neutron absorption of Dy) of Dy$_2$Be$_2$GeO$_7$ could also determine the crystal electric fields, which would definitively determine if Dy$_2$Be$_2$GeO$_7$ is an effective spin $1/2$ system; determine the magnetic structure of the long-range ordering under zero-field conditions; and probe the development of the ostensible short-range correlations with field.

\section{Conclusion}
We have shown that Dy$_2$Ge$_2$O$_7$ and Gd$_2$Ge$_2$O$_7$ are antiferromagnetic below T$_N$ $\sim$ 1~K. Dy$_2$Ge$_2$O$_7$ is consistent with an Ising-like antiferromagnet, having a powder saturation magnetization that reaches about half (5.4(1)~$\mu_B/$Dy$^{3+}$) the expected free moment and a nearly zero isothermal susceptibility at low fields. The long-range antiferromagnetic order in Dy$_2$Ge$_2$O$_7$ is replaced by a spin-flip phase in fields greater than 86(1)~mT at 500~mK. Conversely, Gd$_2$Ge$_2$O$_7$ is isotropic, having a saturation magnetization that reaches the full free-ion moment. A quadratic contribution to the magnetization is observed in the ordered state, and a metamagnetic transition is detected for a field of 790(10)~mT at 500~mK. This transition may be associated with dipole-dipole interactions, or the Dyzaloshinskii-Moriya interaction.

\section{Acknowledgements}

We thank Dr. Bruce Gaulin and Dr. Erik S. S{\o}renson for their helpful discussions. Work at McMaster University was supported by the Natural Sciences and Engineering Research Council of Canada.

\end{document}